\documentclass[aip, jap, reprint, showkeys, showpacs]{revtex4-1}

\usepackage{graphicx}
\usepackage{amsmath}
\usepackage[latin1]{inputenc}
\usepackage[usenames,dvipsnames]{color}
\makeatletter
\DeclareRobustCommand*\textsubscript[1]{%
  \@textsubscript{\selectfont#1}}
\def\@textsubscript#1{%
  {\m@th\ensuremath{_{\mbox{\fontsize\sf@size\z@#1}}}}}
\makeatother

\begin{document}

\title{Structural recovery of ion implanted ZnO nanowires}

\author{G. Perillat-Merceroz}
\email{gperillat@gmail.com}
\affiliation{CEA, LETI, Minatec Campus, Grenoble, 38054, France}
\affiliation{CEA INAC/UJF-Grenoble1 UMR-E, SP2M, LEMMA, Minatec
Campus, Grenoble, 38054, France}
\author{F. Donatini}
\affiliation{Institut Néel, CNRS and Université Joseph Fourier, BP 166, 38042 Grenoble Cedex 9, France}
\author{R. Thierry}
\affiliation{CEA, LETI, Minatec Campus, Grenoble, 38054, France}
\author{P.-H. Jouneau}
\affiliation{CEA INAC/UJF-Grenoble1 UMR-E, SP2M, LEMMA, Minatec
Campus, Grenoble, 38054, France}
\author{P. Ferret}
\affiliation{CEA, LETI, Minatec Campus, Grenoble, 38054, France}
\author{G. Feuillet}
\affiliation{CEA, LETI, Minatec Campus, Grenoble, 38054, France}

\begin{abstract}
Ion implantation is an interesting method to dope semiconducting materials
such as zinc oxide provided that the implantation-induced defects
can be subsequently removed. Nitrogen implantation followed by anneals
under O\textsubscript{2} were carried out on zinc oxide nanowires
in the same conditions as in a previous study on bulk ZnO {[}J. Appl.
Phys. 109, 023513 (2011){]}, allowing a direct comparison of the defect
recovery mechanisms. Transmission electron microscopy and cathodoluminescence
were carried out to assess the effects of nitrogen implantation and
of subsequent anneals on the structural and optical properties
of ZnO nanowires. Defect recovery is shown to be more effective in
nanowires compared with bulk material due to the proximity of free
surfaces. Nevertheless, the optical emission of implanted and annealed
nanowires deteriorated compared to as-grown nanowires, as also observed
for unimplanted and annealed nanowires. This is tentatively attributed
to the dissociation of excitons in the space charge region induced
by O\textsubscript{2} adsorption on the nanowire surface.
\end{abstract}

\maketitle


\section{Introduction}

For a decade, zinc oxide has aroused growing interest, especially
because of potential applications in short-wavelength optoelectronics.
This material is particularly interesting as nanowires, which are
easily grown, with no catalysts, on various types of substrates such
as sapphire, and by different growth methods such as metalorganic
vapor phase epitaxy (MOVPE).\cite{park_metalorganic_2002} Among other
potential applications, ZnO nanowires are studied for making light-emitting
diodes (LEDs) because of the advantages they present over ZnO thin
layers, related for instance to the presence of larger developed surfaces.
Nanowire growth on large and conductive hetero-substrates such as
silicon\cite{lee_comparative_2005} and metal\cite{park_catalyst-free_2008}
is possible, and no extended defects are expected (for example, none
were observed in nanowires grown on sapphire).\cite{rosina_morphology_2009,perillat-merceroz_mocvd_2010}
Furthermore, nanowires with radial core-shell quantum wells can be
grown, yielding large emitting volumes,\cite{bae_fabrication_2006}
and with a good internal quantum efficiency.\cite{thierry_core-shell_2012}
Moreover, light extraction is naturally more efficient in nanowire
LEDs than in thin layer LEDs.\cite{henneghien_optical_2011} \emph{p}-type
doping of ZnO nanowires remains challenging, but\emph{ p}-type doping
obtained \emph{in situ} during the growth was reported, allowing the
observation of a weak ultra-violet (UV) emission under current injection
for \emph{p-n} homo-junctions in ZnO nanowires.\cite{willander_zinc_2009-1,chen_near_2010}
Finally, an electrically-pumped laser made of \emph{in situ} Sb-doped
\emph{p}-type ZnO nanowires on a \emph{n}-type ZnO thin film was demonstrated.\cite{chu_electrically_2011}

Ion implantation is another way to dope ZnO, which offers the possibility
to introduce dopants beyond their solubility limit. However this method
degrades the optical and electrical properties because of the creation of structural defects,\cite{kucheyev_ion-beam-produced_2003} such as point defects\cite{chen_interaction_2005} and dislocation loops.\cite{perillat-merceroz_formation_2011}
These defects can be removed by annealing, but the temperature must
be sufficiently high to recover a material without defects, and sufficiently
low in order not to deactivate the dopants. Two publications reported
on LEDs made from ZnO nanowires for which \emph{p}-type doping was
tried by ion implantation.\cite{yang_p-n_2008,sun_ultraviolet_2009}
In the first one, \emph{p}-type doping was claimed through arsenic
implantation followed by a 750\textdegree{}C annealing for 2~h, with
doses of 10\textsuperscript{14} or 10\textsuperscript{15}~cm\textsuperscript{-2}.
For the higher 10\textsuperscript{15}~cm\textsuperscript{-2} dose,
UV electro-luminescence (EL) was negligible compared to the red one.
For the lower 10\textsuperscript{14}~cm\textsuperscript{-2} dose,
EL was obtained mainly in the UV range, but the signal was very noisy
and appeared above a rather high voltage (6~V). The second publication
concerns phosphor implantation with a 10\textsuperscript{14}~cm\textsuperscript{-2}
dose. After a 900\textdegree{}C annealing for 2~h, UV EL was also
measured, but once again only for high voltages and currents (a few
tens of V and mA). Moreover, the EL may be due to a metal-insulator-semiconductor
stack. These publications clearly call for an optimization of the
annealing conditions of the implantation-induced defects.

Recovery of structural defects by annealing in ZnO nanowires has been
studied for manganese, vanadium, and gallium implantations.\cite{ronning_manganese-doped_2004,schlenker_properties_2007,yao_structural_2009}
For Mn implantation, it was shown by photoluminescence (PL) that nanowires
recovered nearly the same luminescence as unimplanted ones after a
800\textdegree{}C, 15~min annealing under vacuum.\cite{ronning_manganese-doped_2004}
However, it was shown by transmission electron microscopy (TEM) that
the dislocation loops identified before annealing did not disappear
totally. For V implantation, a partial recovery of implantation defects
after a 500\textdegree{}C, 30~min annealing under O\textsubscript{2}
was observed.\cite{schlenker_properties_2007} Concerning Ga implantation,
a parametric study as a function of the implanted dose (from 5$\times$10\textsuperscript{12}
to 1.5$\times$10\textsuperscript{16} cm\textsuperscript{-2}), and
of the annealing temperature (from 450\textdegree{}C to 700\textdegree{}C)
was carried out.\cite{yao_structural_2009} For doses lower than 5$\times$10\textsuperscript{13}
cm\textsuperscript{-2}, implantation-induced dislocation loops disappeared
after annealing under argon at 700\textdegree{}C.

However, no comparison was made in these studies between ZnO nanowires
and bulk ZnO in terms of implantation defect recovery. Thus it is
not known whether implanting nanowires is advantageous or not compared
to bulk implantation. We previously demonstrated, for the bulk case,
that dislocation loops induced by nitrogen implantation were hard
to remove by annealing, leading to poor electrical properties.\cite{perillat-merceroz_formation_2011}
However, the dislocations were shown to disappear just below the surface
of the implanted substrate, suggesting that if implantation was carried
out in thin enough nanowires, a total structural recovery could be
possible. Moreover, despite numerous works on N implantation of bulk
samples or thin layers, no study combines both TEM and optical spectroscopy
concerning the removal of the nitrogen implantation defects, a crucial
step to limit non-radiative losses in light-emitting devices.

In this paper, N implantation was carried out on nanowires using the
same conditions as in our previous study on bulk implantation.\cite{perillat-merceroz_formation_2011}
This allows a direct comparison of the defect recovery mechanisms.
Annealing at different temperatures and for different times were performed.
TEM and cathodoluminescence (CL) were used to compare spatially-resolved
structural and optical properties of as-grown, implanted, implanted-annealed,
and unimplanted-annealed nanowires. Implantation defect recovery is
shown to be more effective in nanowires compared with bulk material.
Nevertheless, even if dislocation loops disappear after annealing
because of the proximity of the free surfaces, the optical properties
are not recovered compared to the as-grown nanowires. The optical
properties are degraded similarly for unimplanted and annealed nanowires,
although these do not contain any extended defects, possibly because
of O\textsubscript{2} adsorption on the nanowire surfaces.

\section{Experimental details}

Nanowires were grown by MOVPE on sapphire. More details about the
growth conditions and the structural characterizations are presented
in previous articles.\cite{rosina_morphology_2009,perillat-merceroz_mocvd_2010,perillat-merceroz_compared_2012}
Two unimplanted parts of the nanowire sample were kept as references
for TEM and CL studies: one as-grown and the other one annealed at
900\textdegree{}C for 2~h under O\textsubscript{2} at atmospheric
pressure. On another part of the sample, three successive implantations
at different energies and doses were carried out in order to obtain
a flat nitrogen profile. Accelerating voltages of 50, 120, and 190~kV
with respective doses of 4.10\textsuperscript{14}, 8.10\textsuperscript{14},
and 10\textsuperscript{15}~cm\textsuperscript{-2} were used. The
sample was tilted 7\textdegree{}C relative to the ion beam direction.
One piece of the implanted sample was annealed at 700\textdegree{}C
for 15~min, another at 900\textdegree{}C for 15~min and the last
one at 900\textdegree{}C for two hours, under O\textsubscript{2}
at atmospheric pressure. CL experiments were carried out at 30~kV
and 10~K. TEM was done on a FEI-Tecnai microscope operated at 200~kV.

\section{Results}

\subsection{Structural properties}

\begin{figure}
\begin{centering}
\includegraphics[width=8cm]{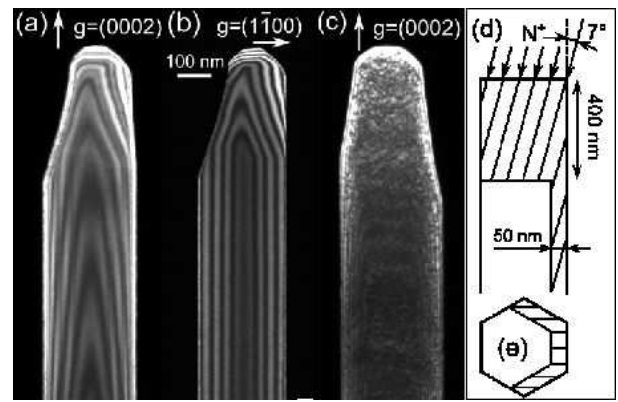}
\par\end{centering}

\caption{Weak-beam TEM images of an as-grown nanowire with (a) $g=\left(0002\right)$,
with (b) $g=\left(1\bar{1}00\right)$, and of (c) an implanted nanowire
with $g=\left(0002\right)$. Schematics of the implantation depth
as simulated with the SRIM software\cite{ziegler_srim_2010} (d) in
cross section, and (e) in a plane section of the nanowire bottom.}

\centering{}\label{Fig1}
\end{figure}

Weak-beam TEM images of as-grown nanowires with $g=\left(0002\right)$,
and with $g=\left(1\bar{1}00\right)$ {[}Fig.~\ref{Fig1}~(a) and
(b) respectively{]} exhibit only equal thickness fringes: neither
stacking faults nor dislocations are present. Implantation-induced
structural defects can be seen in Fig.~\ref{Fig1}~(c). In our previous
work about N implantation in bulk ZnO, the same defects were observed.\cite{perillat-merceroz_formation_2011}
These are dislocation loops formed by agglomeration of Zn and O interstitials.
But contrary to the bulk case, loops are present here not only in
a 400 nm-thick layer below the top surface, but also on the side-walls
of the nanowires because of the inclination of the ion beam {[}Fig.~\ref{Fig1}~(d){]}.
The nanowire density and height (around 5$\times$10\textsuperscript{6}~cm\textsuperscript{-2}
and 3~$\mu m$) are actually such that shadowing effects are negligible.

\begin{figure}
\begin{centering}
\includegraphics[width=8cm]{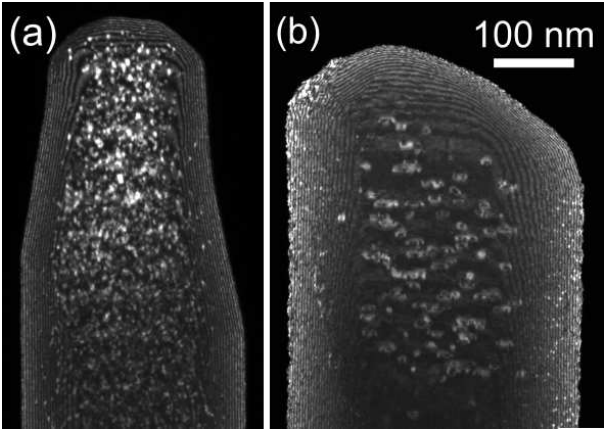}
\par\end{centering}

\caption{Weak-beam TEM images with $g=\left(0002\right)$ of implanted nanowires
after annealing (a) at 700\textdegree{}C for 15~min, and (b) at 900\textdegree{}C
for 15~min.}

\centering{}\label{NF implantes recuits}
\end{figure}

Figure~\ref{NF implantes recuits} shows TEM images of implanted
and annealed nanowires for two annealing conditions. For a 700\textdegree{}C,
15~min annealing{[}Fig.~\ref{NF implantes recuits}~(a){]}, a 35~nm
outer shell with no remaining dislocation loops is observed. Dislocations
are still visible below the 400~nm zone under the top of the nanowire,
because the implanted thickness on the sides is thicker (50~nm) than
the recovered outer shell (35~nm). For a 900\textdegree{}C 15~min
annealing {[}Fig.~\ref{NF implantes recuits}~(b){]} the outer shell
is 70~nm wide, as observed for the sub-surface zone in the bulk case.\cite{perillat-merceroz_formation_2011}
Contrary to the 700\textdegree{}C annealing case, there are no dislocations
below the 400~nm zone under the nanowire top, because the recovered
outer shell is thicker (70~nm) than the implanted shell on the sides
(50~nm) (see also figure \ref{TEM+CL} (a)). As observed for the
bulk case, loops are less dense but bigger with increased annealing,
which is ascribed to a larger point defect diffusion at high temperature.\cite{perillat-merceroz_formation_2011} 

\begin{figure}
\begin{centering}
\includegraphics[width=8cm]{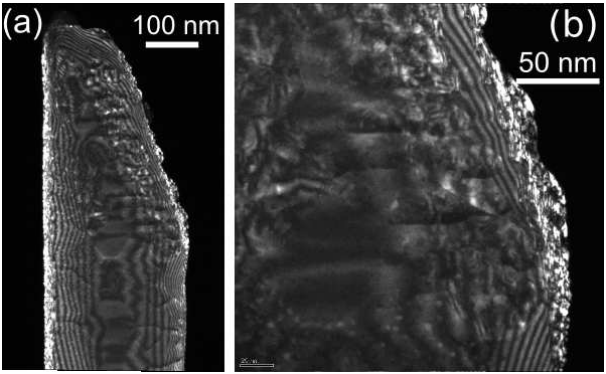}
\par\end{centering}

\caption{Weak-beam TEM images with $g=\left(0002\right)$ of implanted nanowires
after annealing at 900\textdegree{}C for 2~h, at two different magnifications.}

\centering{}\label{NF N 900 2h}
\end{figure}

\begin{figure}
\begin{centering}
\includegraphics[width=8cm]{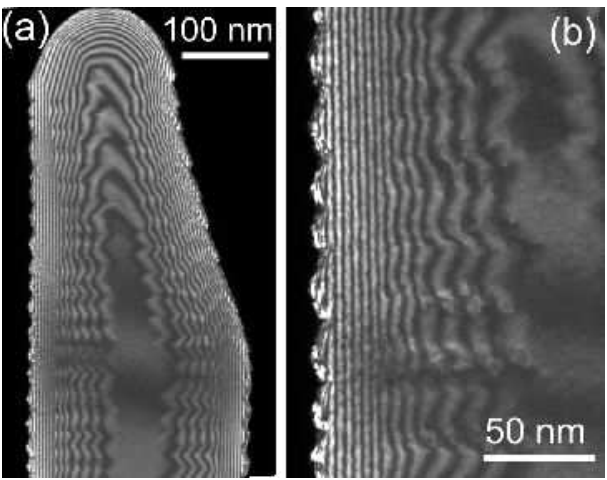}
\par\end{centering}

\caption{Weak-beam TEM images with $g=\left(0002\right)$ of unimplanted nanowires
after annealing at 900\textdegree{}C for 2~h, at two different magnifications.}

\centering{}\label{tem ni}
\end{figure}

Figure~\ref{NF N 900 2h} shows TEM images of a nanowire implanted
and annealed at 900\textdegree{}C for two hours. The nanowire surface
appears rough, and some defects are visible at the edges of the nanowire,
but not in its core. It is thus unlikely that these defects are implantation-induced
dislocation loops, because they should be present in the core too.
Moreover, dislocations were shown to disappear in a 70~nm zone from
the surface after a short 15~min, 900\textdegree{}C annealing. Therefore,
it is not expected that for a longer annealing, dislocations would
be present in this zone. Figure \ref{tem ni} shows an unimplanted
nanowire annealed in the same conditions (900\textdegree{}C, 2~h
under O\textsubscript{2}) for comparison. The nanowire exhibits rough
side facets instead of the perfectly flat ones for as-grown samples.
Despite this, no structural defects are visible inside the nanowire.
Consequently, the defects visible in figure~\ref{NF N 900 2h} are
probably due to an annealing-induced surface degradation of the implanted
ZnO, as this was already observed for high dose arsenic implantation
at a 1.4$\times$10\textsuperscript{17}cm\textsuperscript{-2} dose,
after annealing at 1000\textdegree{}C.\cite{coleman_thermal_2005}
This stresses the fact that annealing conditions have to be optimized:
sufficient annealings have to be performed in order to remove the
implantation defects, but moderate annealings (short duration and
low enough temperature) is desirable in order not to degrade the implanted
ZnO. Moreover, moderate annealings are necessary not to deactivate
the dopants.\cite{fons_direct_2006} For nanowires smaller than 140~nm
or 70~nm in diameter, it is extrapolated that no dislocations should
be left after annealing at 900\textdegree{}C or 700\textdegree{}C
respectively, for a duration of no more than 15~min. It is thus demonstrated
that the elimination of dislocations is facilitated by the proximity
of the nanowire free surfaces, allowing shorter duration annealings
at lower temperatures, which is a clear advantage of nanowires compared
to bulk material.

\subsection{Optical properties}

\begin{figure}
\begin{centering}
\includegraphics[width=8cm]{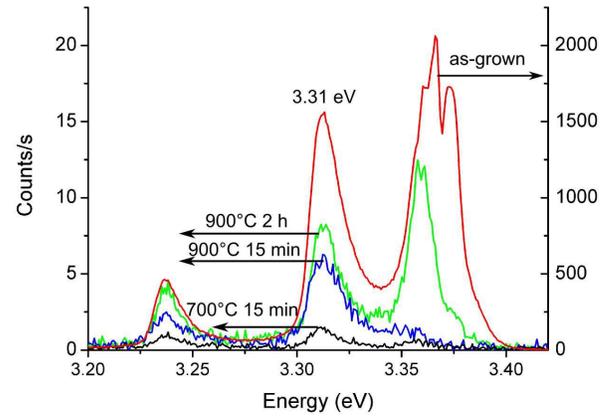}
\par\end{centering}

\caption{10~K CL spectra from the top of an as-grown nanowire (scale on the
right axis), and from the tops of the implanted and annealed nanowires
at 700\textdegree{}C for 15~min, at 900\textdegree{}C for 15~min,
and at 900\textdegree{}C for 2~h (scale on the left axis).}

\centering{}\label{CL NF implantes recuits}
\end{figure}

Figure~\ref{CL NF implantes recuits} shows the CL spectrum taken
on a 400~nm-long zone at the top of an as-grown nanowire (curve in
red, with the scale on the right axis). Apart from the excitonic peaks
around 3.365~eV,\cite{meyer_bound_2004} a peak at 3.31~eV is present,
with its one longitudinal optical phonon replica separated by 72~meV.
The exact origin of the 3.31~eV peak has been debated a lot (see
for example Ref. \cite{schirra_stacking_2008,al-suleiman_mechanisms_2009}),
and is beyond the scope of this paper: we will only discuss the relative
intensities of the peaks depending on the different annealing conditions.
The CL spectrum of an as-grown nanowire is compared to that of the
top part of an implanted and annealed nanowire. Upon implantation
and annealing, a drastic decrease of the emission is observed (by
a factor of 200 for the 3.31~eV peak after annealing at 900\textdegree{}C
for 2~h). For higher temperature or longer annealing, the emission
is slightly improved: considering for instance the 3.31~eV peak,
there is an increase by a factor of three between the emission of
the nanowires annealed at 700\textdegree{}C for 15~min and at 900\textdegree{}C
for 2~h. The 3.36 eV excitonic emission recovers more slowly and
appears only when the density of dislocations is low enough: it is
nearly absent for nanowires implanted and annealed at 900\textdegree{}C
for 15~min, whereas it is present, although weaker than for as-grown
nanowires, for a longer annealing of two hours at 900\textdegree{}C. In the implanted and annealed samples, it is impossible to
detect the presence of the donor acceptor pair transition attributed to N at 3.235 eV,\cite{meyer_bound_2004} because of the phonon replica of the 3.31 eV peak around 3.24 eV.

\begin{figure}
\begin{centering}
\includegraphics[width=8cm]{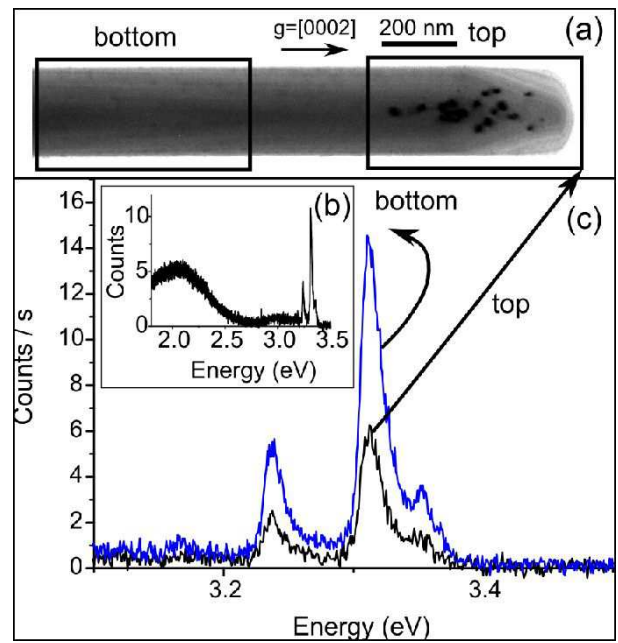}
\par\end{centering}

\caption{Implanted nanowires after annealing at 900\textdegree{}C for 15~min:
(a) bright-field TEM image with $g=\left(0002\right)$ diffracting
conditions showing the top of a nanowire with remaining dislocations
and the bottom free of dislocations, (b) CL spectrum of a whole nanowire
including the deep defect band, and (c) 10~K CL spectra taken at
the top and at the bottom of a nanowire.}

\centering{}\label{TEM+CL}
\end{figure}

Figure~\ref{TEM+CL}~(a) shows a TEM image of an implanted nanowire
after annealing at 900\textdegree{}C for 15~min, at a lower magnification
than in Fig.~\ref{NF implantes recuits}~(b). The black dots reveal
the presence of dislocation loops in the core of the top part of the
nanowire while, in the the bottom part, no dislocations are visible.
A CL spectrum taken on another nanowire but from the same sample reveals
the presence of a deep defect band between 1.5 and 2.5~eV {[}Fig.~\ref{TEM+CL}~(b){]}
which, as shown in figure 7, is very weak in as-grown nanowires relatively to the near-band edge emission (about a hundred times less).\cite{robin_evidence_2007}
Moreover, CL spectra are acquired at the bottom and at the top of
the nanowire {[}Fig.~\ref{TEM+CL}~(c){]}. The 3.31~eV emission
is stronger in the bottom part of the wire, with no more dislocations
(14~counts/s), than at the top (6~counts/s). However it remains
very weak compared to the emission of as-grown nanowires (1500~counts/s,
see Fig. \ref{CL NF implantes recuits}): although implantation-induced
structural defects are removed, the optical properties remain very
poor. In order to have a better understanding of the reasons behind
this weak optical emission, the optical properties of unimplanted
and annealed nanowires were also examined.

\begin{figure}
\begin{centering}
\includegraphics[width=8cm]{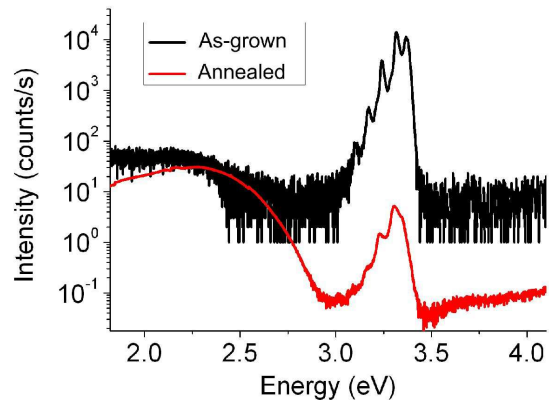}
\par\end{centering}

\caption{CL spectra at 10~K taken at the top of an unimplanted nanowire (as-grown,
in black), and of unimplanted and annealed at 900\textdegree{}C for
2~h nanowire (in red).}

\centering{}\label{CL ni}
\end{figure}

Figure \ref{CL ni} compares the CL spectra of an as-grown nanowire,
and of a 900\textdegree{}C, 2~h annealed nanowire. Before annealing,
and as pointed out earlier, the visible band to near-band-edge intensity
ratio is about 10\textsuperscript{-2} whereas it reaches about 10
after annealing. Interestingly, the visible emission intensity is
about the same before and after annealing, while the near-band-edge
emission alone is drastically reduced upon O\textsubscript{2} annealing.
Therefore, since Zn vacancies and O vacancies are known to contribute
to the visible emission (at 1.6~eV for zinc vacancy clusters, 1.9-2.1~eV
for zinc vacancies, and 2.3~eV for O vacancies),\cite{dong_vacancy_2010}
the results from figure \ref{CL ni} suggest that annealing under
O\textsubscript{2} does not change the vacancy concentrations much.
On the other hand, in order to explain the loss in near-band-edge
emission, extended defects are ruled out, since they are not observed
in unimplanted-annealed nanowires (see Fig.\ref{tem ni}). Furthermore,
extended defects are non radiative defects, which woul\textcolor{black}{d
lead to an overall decrease of the PL intensity on the whole spectrum.
The origin of the weaker near-band-edge emission upon annealing, and
the correlated constant visible emission, is yet unclear. Some mechanisms
have been proposed to account for the lower luminescence because of
an O\textsubscript{2} atmosphere. Photo-current measurements of ZnO
nanowires under UV illumination were shown to depend upon the ambient
atmosphere (wet or dry air, or vacuum).\cite{fan_photoluminescence_2004,li_competitive_2009,sohn_influence_2009}
This was attributed to the presence or absence of adsorbed O\textsubscript{2}
molecules which would act as electron traps, leading to the formation
of O\textsubscript{2}\textsuperscript{-}. Consequently, excitons
are easily dissociated by the electric field in the created space
charge region, while transitions involving deep levels in the gap
(as is the case for the visible emission) would be much less affected.}

In the literature about implanted and annealed nanowires, various
cases were reported concerning the optical properties. For Ga implantation
with doses similar to ours, annealing under argon did not change the
optical emission.\cite{yao_structural_2009} For V implantation, the
emission of implanted nanowires was improved after annealing under
O\textsubscript{2}, as observed in our case, but the PL intensity
was not compared to the one before implantation.\cite{schlenker_properties_2007}
Finally, for Mn implantation of ZnO nano-ribbons, it was noticed that
the emission of nanowires implanted and annealed at 800\textdegree{}C
under vacuum conditions had an emission nearly as high as before implantation,
together with a lower density of structural defects.\cite{ronning_manganese-doped_2004}
Although no consensus can be drawn from these observations, we point
out that annealing under vacuum might not be detrimental to the optical
properties, contrary to annealing under O\textsubscript{2}.

\section{Conclusion}

To sum up, it was found that implantation-induced dislocation loops
were removed more easily by annealing in ZnO nanowires than in ZnO
bulk samples thanks to the proximity of free surfaces. This important
result may be generalized to nanowires of other materials. For nanowires
with small enough diameters, dislocation loops can be totally eliminated
at lower temperatures than for bulk samples. This is interesting because
high temperature anneals may deactivate the dopants and, as we have
shown here, may also lead to a degradation of the structure at the
surface. However, post-implantation anneals under O\textsubscript{2}
were found to alter the optical properties of ZnO nanowires especially
in the near-band-edge region of the spectrum. We tentatively attributed
this behavior to the space charge region at the free surfaces induced
by O\textsubscript{2} adsorption. To conclude on ZnO doping by ion implantation, it seems difficult because optical properties are not recovered after annealing.

\section*{Acknowledgments}

The authors acknowledge C. Granier and M. Lafossas for their technical
assistance, J.-P. Barnes for making some corrections to the English of this publication, and funding from the French national research agency (ANR)
through the Carnot program (2006/2010).

\bibliography{biblio}

\end{document}